# Contribution of Glottal Waveform in Speech Emotion: A Comparative Pairwise Investigation

Zhongzhe Xiao, *Member, IEEE*, Ying Chen, Zhi Tao

School of Optoelectronic Science and Engineering

Soochow University

Suzhou, Jiangsu Province, P.R.China

Email: xiaozhongzhe@suda.edu.cn, taoz@suda.edu.cn

*Abstract*—In this work, we investigated the contribution of the glottal waveform in human vocal emotion expressing. Seven emotional states including moderate and intense versions of three emotional families as anger, joy, and sadness, plus a neutral state are considered, with speech samples in Mandarin Chinese. The glottal waveform extracted from speech samples of different emotion states are first analyzed in both time domain and frequency domain to discover their differences. Comparative emotion classifications are then taken out based on features extracted from original whole speech signal and only glottal wave signal. In experiments of generation of a performance-driven hierarchical classifier architecture, and pairwise classification on individual emotional states, the low difference between accuracies obtained from speech signal and glottal signal proved that a majority of emotional cues in speech could be conveyed through glottal waveform. The best distinguishable emotional pair by glottal waveform is intense anger against moderate sadness, with the accuracy of 92.45%. It is also concluded in this work that glottal waveform represent better valence cues than arousal cues of emotion.

## I. INTRODUCTION

Speech is one of the most important way for human to express emotion in their vocal communications. Research on vocal emotion has become a topic of interest in the field of affective computing [1]. One of the development direction of speech emotion is to introduce multi-modal signals into the study of emotion, which has achieved remarkable results in recent years, including facial images and/or video clips [2] [3] [4] [5] [6] [7] [8], or physiological signals [9]. This is an important study mode of emotion, because people express their emotions in an integrated way through speech, facial expression, body gesture, etc. But as there still exists such situations with no cameras or poorly illuminated, that signals cannot be conveniently collected, except speech. Thus, research on emotion with only speech signal is still an essential topic in affective computing.

Another approach with interest of emotional speech is the cross-culture study [10] [11] [12] [13] [14]. Actually, in order to find out the commonness of speech emotion expressing manner from multiple cultures, a prospective way is to start from the physiological base of human voicing process, i.e. glottal source, and vocal tract filtering. We will focus in this paper on the contribution of glottal waveform in emotion expressing.

The stimulation source of human speech, glottal waveform, can be reflected by electroglottograph (EGG), which can be collected by attaching sensors on the skin near glottis with a special equipment. It has been proved that EGG signals can exhibit speech emotions [9], thus the glottal waveform should also be able carry emotional information of speech. However, EGG signals are not collectible in natural speaking in real life, without the sensors staying on the neck. Fortunately, researcher started the studies on the glottal waveform [15] and the relationship between EGG and glottal waveform [16] [17] [18] from a very early stage. Currently there are already a number of reliable method to obtain glottal waveform from speech, though inverse filtering [19] [20] [21] [22] [23].

With the glottal signal extracted from speech, several studies on emotional speech using glottal features have been made on positive emotions such as stress or depression for clinic purpose [24] [25] [26]. Emotional types that are more common in daily life, e.g. anger, joy, sadness, are considered in this work. Similar to [27], we make pairwise studies in emotional pairs from 7 emotional states, while instead of evaluating only the glottal parameters, comparative experiments are implemented in this work, by comparison of the performance in emotion distinguishing ability with speech signal and with glottal signal. Thus, we can find out in what degree that glottal waveform could convey emotional cues that are contained in speech. The lower the difference, the higher the emotion expressing ability by glottal waveform.

The paper is organized as follows. The overall method is described in Section 2. Section 3 introduces the inverse filtering method adopted in this work, and a brief analysis of glottal waveform from speech samples of different emotional states. Section 4 gives out the experiments on comparative classifications and results. Conclusions are drawn in Section 5.

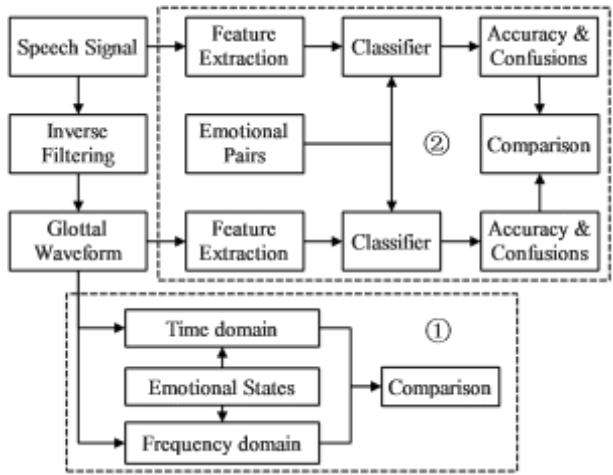

Figure 1. The comparative method in investigating the contribution of glottal wave in vocal emotion expressing.

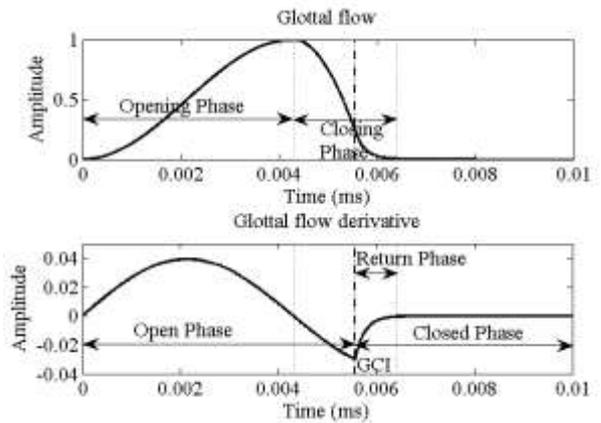

Figure 2. The glottal flow waveform and the glottal flow derivative waveform. Figure adapted from [22].

## II. METHOD

Acting as the simulation source of speech signal in source-filter models, glottal waveform is an essential part in all aspects of human vocal expressing, including emotion expressing. In this work, we investigated the contribution of glottal wave in human vocal emotion expressing in a comparative way.

We first separate the glottal waveform from the speech signal by means of inverse filtering. Then two analyses are performed, as shown in the two dotted frames of Fig. 1.

The first analysis is an intuitive one, that is, to compare the glottal waveform from the different emotional states. The samples of glottal waveform are compared mainly within emotional families, with the neutral state as reference. In time domain, the waveform shapes of increase in the opening phase and the decrease in the closing phase are compared. In frequency domain, the fundamental frequency and its bandwidth, and the amplitude & bandwidth of the second harmonic, are the most representative parameters for emotions. This analysis will be stated in Section 3.

The second analysis is a machine learning base one. The original whole speech signal and the extracted glottal waveform are treated as two types of data sources. For each type of the data source, speech signal or glottal waveform, features are extracted on the same feature set definition, and fed into classifiers with the same settings for training. The classification results, including accuracies and confusion matrices are compared for analysis. Note that the glottal waveform is only the stimulation source part of speech signal. Another important factor in speech generating and emotion expressing, the vocal tract filtering effect, is eliminated. That is to say, some of the emotional cues in speech are not included in glottal waveform. So, it is normal that the classification results from glottal waveform are lower than those from the whole speech signal. If the difference between the results from the two types of data source is low enough, we could verify that the stimulation part of speech (glottal waveform) is able to carry most emotional information of speech. This analysis will be stated in Section 4.

## III. GLOTTAL WAVEFORM OF EMOTIONAL SPEECH

In this section, we first introduce the extraction process of glottal waveform from emotional speech samples. Then, we make a brief analysis to the glottal waveform of different emotional states

### A. Glottal Inverse Filtering

In the source-filter model, the speech generation can be defined in z-domain as:

$$S(z) = G(z)V(z)L(z) \qquad (1)$$

where $S(z)$ is the speech signal, $G(z)$ is the glottal flow waveform, which corresponds to the stimulate source, and $V(z)$ and $L(z)$ correspond to the filter, which present the vocal tract transfer function and the lip radiation effect respectively, to convert the air volume velocity waveform into an acoustic pressure wave signal - speech.

The stimulation source, *i.e.* the glottal flow waveform, is an air flow streaming from the lungs pushing the vocal folds to oscillate. The vibration of the vocal folds cycles from an open position into a closed position, and generates the open phase and closed phase in glottal flow pulses. The border of the open phase and closed phase is an abrupt stop of air flow that is called the glottal closure instant (GCI), which is obvious in the glottal flow derivative waveform. The glottal flow and the glottal flow derivative following the classic LF model [15] are illustrated in Fig. 2.

To extract glottal waveform from speech signals, which can be done with glottal inverse filtering (GIF), is an important way to understand the process of human voicing. Researchers have proposed a number of known GIF algorithms over a wide range of time, e.g. closed phase (CP) covariance technique [19], iterative adaptive inverse filtering (IAIF) [20], complex cepstral decomposition (CCD) [21].

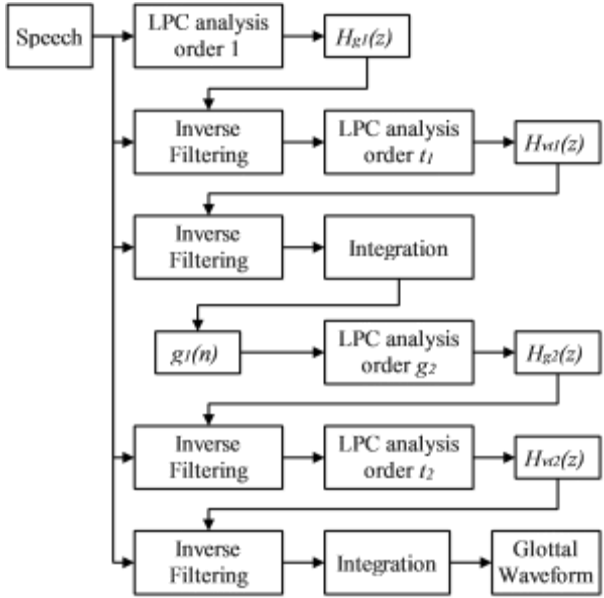

Figure 3. IAIF method of glottal waveform extraction

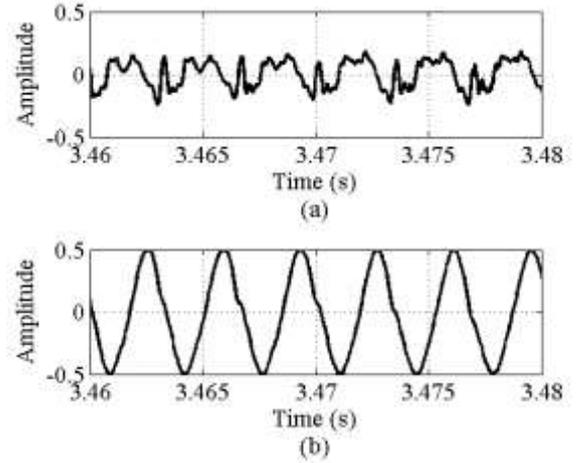

Figure 4. A sample of speech signal and its glottal waveform extracted with IAIF. (a) Speech signal (b) Glottal waveform.

IAIF method is adopted in the glottal wave extraction in this work. Although among the early GIF algorithms, IAIF is still among the most precise algorithms, and is still the base of several currently used GIF method [22] [23]. The IAIF includes two stages, as shown in Fig. 3.

The first stage begins with a first order LPC, to eliminate the lip radiation effect, as Hg1(z), which reflect the L(z) in eq. 1:

$$H_{g1}(z) = 1 + \alpha z^{-1}, \alpha < 1 \qquad (2)$$

Where α is a parameter in [0.96,1). The speech signal is fed into this filter for the first inverse filtering. Then, by another LPC analysis of order $t_1$, a rough vocal tract filter is obtained as $H_{vt1}(z)$:

$$H_{vt1}(z) = 1 + \sum_{k=1}^{t_1} b(k)z^{-k} \qquad (3)$$

A second inverse filtering is then performed by feeding speech signal into $H_{vt1}(z)$, and integrate to a coarse glottal waveform $g_1(n)$, as the output of the first stage.

In the second stage, an LPC analysis of order g2 is performed to estimate the property of the coarse glottal waveform extracted from the previous stage, and result into a filter $H_{g2}(z)$, which can eliminate the glottal influence from the speech by inverse filtering the speech signal with $H_{g2}(z)$:

$$H_{g2}(z) = 1 + \sum_{k=1}^{g_2} c(k)z^{-k} \qquad (4)$$

The output of this reverse filtering can now be seen as to only reflecting the vocal tract, and thus, another LPC analysis of order $t_2$ can achieve vocal tract filter $H_{vt2}(z)$, which is more accurate than $H_{vt1}(z)$:

$$H_{vt2}(z) = 1 + \sum_{k=1}^{t_2} d(k)z^{-k} \qquad (5)$$

The speech signal is then fed into this final vocal tract filter, and integrated into the desired glottal waveform, $g(n)$. The orders $t_1, g_2, t_2$ are determined upon experiments, and the coefficients $b(k), c(k), d(k)$ are obtained from the LPCs.

The glottal waveform extraction is performed only on voiced part of speech, and adjusted to amplitude range of [0, 1] for analyzing. When extracting features for classification in the next section, the average of glottal waveform is first removed in preprocessing. The glottal waveform of unvoiced part of speech are defined as 0. An example of glottal waveform extracted form a speech sample is illustrated in Fig. 4.

B. Emotion Influence to Glottal Waveform

The speech sample analyzed in this work come from a new Mandarin emotional speech dataset, MES-P (Mandarin Emotional Speech dataset - Portrayed) [28], which is constructed by Soochow University. Three emotional families are considered in this dataset, as anger, joy, and sadness, with a moderate state and an intense state in each family. Together with a neutral state, there are totally 7 emotional states in this dataset. The position of the 7 emotional states in valence-arousal space [29] according to subjective rating of the speech samples are displayed in Fig. 5.

1)  Analysis of glottal waveform in time domain

The first analysis is on the time domain, as shown in Fig. 6. In order to make a clearer view, the time axes of glottal waveform are normalized according to their periods, 4 cycles are displayed for each case. In generating this figure, speech samples from all the emotional states are from the same speaker, on the same sentence in a predefined script, from the same corresponding syllables. The glottal waveforms from the three emotional families are displayed in sub-figures (a), (b), and (c) respectively, dashed line for moderate states, and solid line for intense states. The glottal waveform of neutral state (dashed-dotted line) appears in each sub-figure for reference purpose.

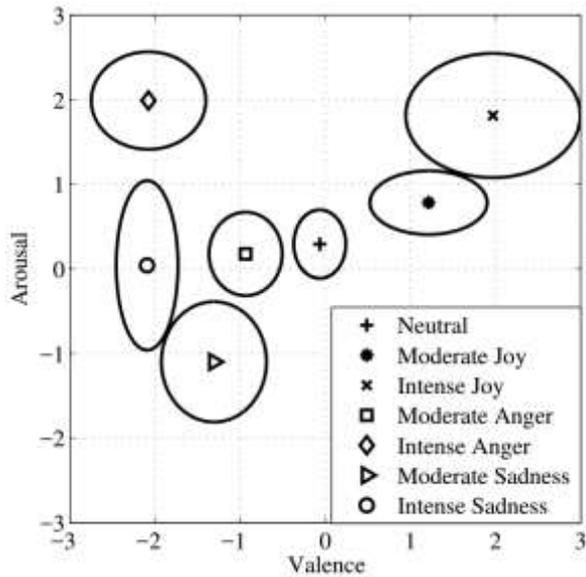

Figure 5. Position of the emotional states in VA space

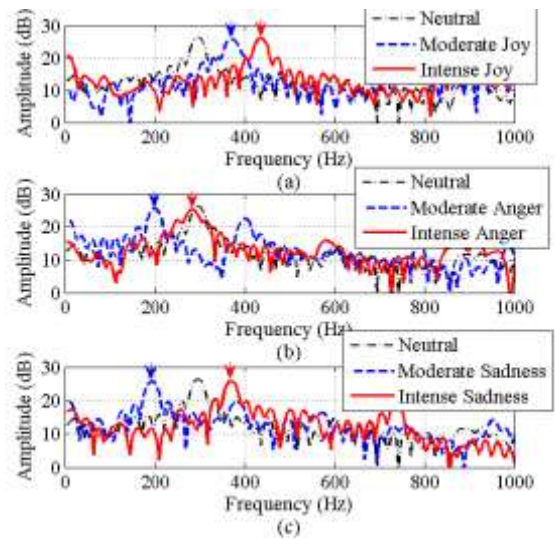

Figure 7. Comparison of glottal waveforms in different emotional states, frequency domain. (a) Joy family (b) Anger family (c) Sadness family.

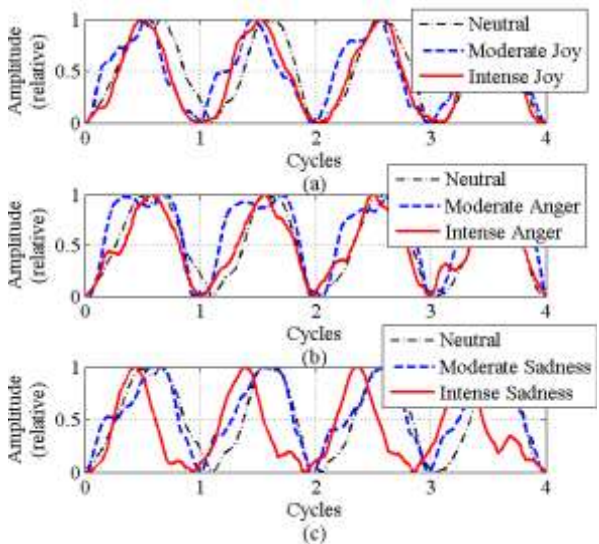

Figure 6. Comparison of glottal waveforms in different emotional states, time domain, normalized in time according to their periods. (a) Joy family (b) Anger family (c) Sadness family.

For joy family (Fig 6 (a)), we can see that the moderate joy presents earlier increase than neutral and intense joy in the opening phase, and both states present earlier decrease than neutral in the closing phase. For anger family (Fig 6 (b)), the moderate anger, or cold anger, presents a wider shape, with an earlier increase in opening phase and later decrease in closing phase, while the intense anger shows similar waveform shape to neutral but with a fluctuation at the beginning of the opening phase. For the sadness family (Fig 6 (c)), the increase in the opening phase of moderate state is a bit earlier than neutral, while the decrease in the closing phase almost coincides neutral. On the contrary, the intense states, has a sharp increase in the opening phase, and its decrease in the closing phase is significantly earlier. Each emotional state exhibits its unique glottal waveform shape.

### 2) Analysis of glottal waveform in frequency domain

The spectrum of glottal waveform from different emotional states is also analyzed. Similar to the case in time domain, we also put the spectrum from the same emotional family in the same sub-figure, with the neutral state for reference, in Fig. 7. The positions of the fundamental frequency ($F0$) of all the emotional states except neutral are marked with arrows in the figure.

In Fig. 7, we can find that the joy emotions exhibit higher $F0$s than neutral, with intense joy even higher; while anger emotions, which also have high arousal in terms of valence - arousal description, gives lower $F0$ for its moderate state, and a very close $F0$ to neutral (still slightly lower) for its intense state. Both anger and joy families show similar pitch tendency within emotional family. The sadness emotions, however, gives out a lower $F0$ for its moderate state (close to that of moderate anger), and a higher $F0$ for its intense state on the contrary (close to that of moderate joy). That is to say, the moderate and intense states of the sadness family could be totally different in the expressing manner, at least for the speaker who provided these speech samples.

In order to make more detailed analysis to the frequency domain glottal waveform, we normalized the spectrum according to their $F0$s in the frequency, and normalized their maximal amplitude to 1, as shown in Fig. 8. We concentrate on two points, the $F0$s and the second harmonic.

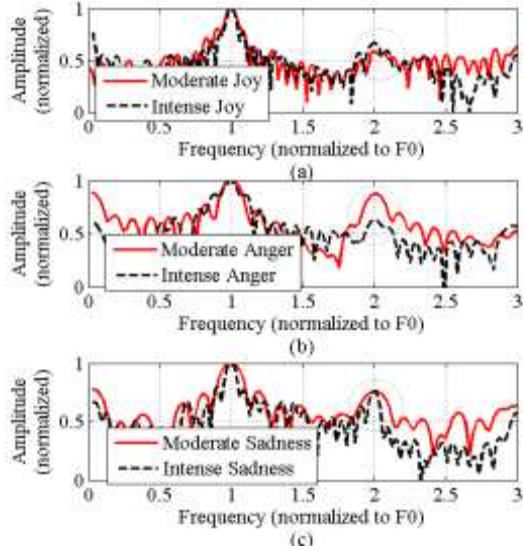

Figure 8. Comparison of glottal waveforms in different emotional states, frequency domain, frequency normalized to F0, amplitude normalized to 1.

The bandwidths near $F0$s for joy emotions (Fig. 8 (a)) are relatively narrow, and those for anger emotions are wider. For sadness emotions, the bandwidths near F0s again show different properties for different intensities, narrower for intense state, and wider for moderate stats.

The second harmonics (marked with dotted ovals) present relatively wider bandwidth than fundamental waves for joy and anger emotional families, and obvious changing in amplitude upon emotional intensities, for joy, intense state is higher, while for anger, moderate state is higher in contrast. For sadness, the amplitudes of the second harmonics are almost the same for moderate and intense states, while the moderate state has a wider bandwidth as in the case near F0s.

The glottal waveform exhibits a various of differences among the 7 emotional states concerned in MES-P dataset, both in time domain and frequency domain. As the sample of glottal waveform illustrated in this subsection come from only one speaker, and only the most basic analysis are carried out, we will show in the next subsection the properties of some of the features which can be extracted from the glottal waveforms, to discover some common regular patterns.

*3) Comparison of certain features extracted from glottal waveform on emotional states pairs.*

In this subsection, we make a primary analysis on the features extracted from the glottal waveforms on the entire MES-P dataset, with 768 samples for each of the 7 emotional states, resulting into totally 5376 samples. The features come from the standard feature set of INTERSpeech 2010 paralinguistic challenge [30] with 1582 features, which will be described in section 4.1.

The analysis is taken out on emotional states pairs, e.g. neutral vs. intense anger, or moderate joy vs. intense sadness, totally 21 possible pairs on the 7 emotional states. The features are ranked in their capacity of distinguishing the two emotional states in each pair, by means of a Bayes-like classification, in a leave-one-speaker-out manner.

Suppose an emotional pair contains two emotional states A and B, there are 768 speech samples in each of the states, from 16 speakers. Speech samples from a certain speaker $T$, 96 samples in the two states, are left as the test set, and the rest of speech samples are used to generate distributions of any feature $F_i$ for the two states. This distribution is seen as the probability density function of feature $F_i$ with speaker $T$ left out, $P_{A_{i_T}}(f_i)$ for emotional state A, and $P_{B_{i_T}}(fi)$ for emotional state B, where the fi is the possible value of feature $F_i$, ranged in $[0, 1]$ for all the features upon a normalization process. Then, for the jth speech sample in the test set, from speaker $T$, we can make a prediction with only feature $F_i$, with its value $f_{ij}$ as:

$$\begin{cases} A, if\ P_{A_{i_T}}(f_{ij}) \geq P_{B_{i_T}}(f_{ij}) \\ B, if\ P_{A_{i_T}}(f_{ij}) < P_{B_{i_T}}(f_{ij}) \end{cases}, j = 1, \cdots, 96 \qquad (6)$$

This process is repeated for speakers $T = 1, \cdots, 16$, thus all the speech samples can get a prediction by single feature $F_i$ on this emotional pair, and result into an accuracy $Accu_i$. The features can then be sequenced with their accuracies from highest to lowest.

The highest accuracies of single feature on the 21 possible emotional state pairs are listed in Table. 1. The best case appears in the pair between intense joy and moderate anger, with an accuracy up to 84.70%, and there are also other 4 pairs with accuracy above 80% (marked with bold font). The worst case appears in the pair between neutral and moderate anger, with an accuracy as low as only 59.64%, slightly above the chance level of 50%. Over all 21 pairs, the average accuracy with a best single feature is 72.27%, and the standard deviation is 7.6%. This phenomenon indicates two possible explanations. First, a certain emotional state may be quite similar or very different to another emotional state; second, the importance of emotional cues expressed in glottal stimulation source depends on the specific emotional states.

For the best pair (intense joy vs. moderate anger) and the worst pair (neutral vs. moderate anger), we also display the properties of the best single features for them, in Fig. 9. Fig. 9 (a) shows the distribution of the best feature for the best pair, solid line for intense joy, and dashed line for moderate anger. These two curves separate from each other with only a small overlapping, thus in Fig. 9 (b) (dots for moderate anger, circles for intense joy), the scattered diagram of this pair with the best two features, a border appears between the majority of points of each state, with a small part cross over each other around the border. Feature $x$ is "F0final_sma_percentile99.0" in the feature set, and feature $y$ is "mfcc_sma[2]_stddev" for this pair, where the feature names are as defined in OPENSmile configuration file. Fig. 9 (c) shows the distribution of the best feature for worst pair, solid line for neutral, and dashed line for moderate anger. The areas covered by the two curves are almost overlapped together, thus in the scatted diagram in Fig. 9 (d) (dots for moderate anger, circles for neutral), the points from the two emotional states are also overlapped without any obvious border, leading to low distinguishably. Feature $x$ is "voicingFinalUnclipped_sma_de_iqr1-2", and feature $y$ is "voicingFinalUnclipped_sma_de_quartile1" for this pair.

TABLE 1. HIGHEST ACCURACIES FOR THE EMOTIONAL PAIRS WITH SINGLE FEATURE (%). "N" STANDS FOR NEUTRAL, "M" STANDS FOR MODERATE, "I" STANDS FOR INTENSE, "J", "A", AND "S" SSAND FOR JOY, ANGER, AND SADNESS RESPECTIVELY.

|     | M-J   | I-J   | M-A   | I-A   | M-S   | I-S   |
| --- | ----- | ----- | ----- | ----- | ----- | ----- |
| N   | 69.04 | **83.92** | 59.64 | **80.73** | 63.87 | 69.14 |
| M-J | -     | 66.93 | 70.90 | 69.08 | 73.57 | 73.57 |
| I-J |       | -     | **84.70** | 62.43 | **84.51** | 72.33 |
| M-A |       |       | -     | 79.62 | 63.36 | 66.99 |
| I-A |       |       |       | -     | 81.51 | 72.66 |
| M-S |       |       |       |       | -     | 68.75 |

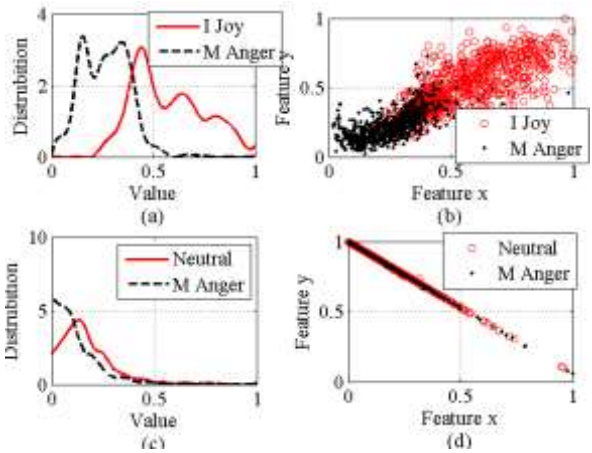

Figure 9. Comparison of distribution of some features on certain emotional states pairs. Letter "I" stands for intense states, and "M" stands for moderate sates. Sub-figures on the left column ((a) (c)) are distribution of the best feature for the pair, sub-figures on the right column ((b) (d))are the scatter diagrams of the best two features for the pair. Emotional pairs: (a) & (b) Intense joy vs. Moderate anger (c) & (d) Neutral vs. Moderate anger. The best features for different pairs are different.

From the above analysis, we assume that although the vocal tract effect has been eliminated from the speech signal in obtaining the glottal waveforms, the features extracted from the glottal waveforms still possess significant emotion discrimination capacity, at least for some of the emotional states, even with selected single feature.

TABLE 2. DESCRIPTORS AND FUNCTIONS IN INTERSPEECH 2010 FEATURE SET. ABBREVIATIONS: DDP: DIFFERENCE OF DIFFERENCE OF PERIODS, ISP: LINE SPECTRAL PAIRS, Q/A: QUADRATIC, ABSOLUTE. TABLE ADOPTED FROM [30].

| Descriptors | Functionals |
| --- | --- |
| PCM loudness | position max./min. |
| MFCC [0-14] | arith. mean, std. deviation |
| log Mel Freq. Band [0-7] | skewness, kurtosis |
| LSP Frequency [0-7] | lin. regression coeff. 1/2 |
| F0 by Sub-Harmonic Sum. | lin. regression error Q/A |
| F0 Envelope | quartile 1/2/3 |
| Voicing Probability | quartile range 2-1/3-2/3-1 |
| Jitter local | percentile 1/99 |
| Jitter DDP | percentile range 99-1 |
| Shimmer local | up-level time 75/90 |

## IV. CLASSIFICATIONS AND RESULTS

In this section, pairwise investigation of contribution of glottal waveform in vocal emotion expression is performed by means of automatic classification. The investigation is taken out in a comparative way, to analysis the difference in performance between two types of sources as original whole speech signal and extracted glottal waveform. Classifications are taken out separately on these two types of sources. The performances from glottal waveform are foreseeable to be not as good as those from whole speech signal, due to the lack of emotional cues generated by the vocal tract, but we expect the differences to be small, so that the glottal waveform could convey a large proportion of vocal emotional cues.

The classifiers adopted for all the predictive models in this work are SVMs, with the SMO (Sequential Minimal Optimization) function, using polynomial kernel. The classifications are performed in leave-one-speaker-out manner.

### A. Feature Set

The standard feature set of INTERSPEECH 2010 paralinguistic challenge (with three sub-challenges as age sub-challenge, gender sub-challenge, and affect sub-challenge) [30] is adopted in this work. The features in this set are extracted from 38 low-level descriptors (LLD) and their first order regression coefficients, with 21 functions applied to these LLDs and their regression coefficients. 16 zero information features are discarded, and finally results into 1582 static features. The descriptors and the functions are listed in Table. 2. The features are extracted using TUM's open-source openSMILE feature extractor [31], with the configuration IS10_paraling.conf. The features are normalized to [0, 1] before sending to classifiers.

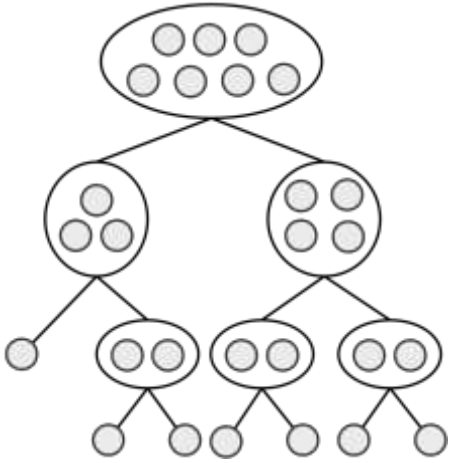

Figure 10. Illustration of the basic hierarchical classifier to be generated.

## B. Experiment I: a Hierarchical Classification Architecture

In this first experiment, we generate a hierarchical classifier architecture as a binary tree, in order to find out the similarity of different emotional states, and the contribution of glottal waveform to distinguish emotional states groups.

To avoid excessive computational burden, we limit the hierarchical in an almost balanced way, that is to say, for the 7 emotional states in consideration, the two emotional states groups in the top level contain 3 states in the left branch, and 4 states in the right branch. The left branch develops into two sub-branches with one or two states in the second level, and the right branch develops into two sub-branches both with two states. All the individual emotional states are separated in the third level (the bottom level). The illustration of the proposed structure is shown in Fig. 10.

### 1) Top level

The concrete hierarchical structure, i.e. the emotional states contained in each branch in the binary tree, is determined by the accuracy of classification result, in a performance-driven way. For the top level, $C_7^3 = 35$ predictors are trained for each of the two types of data sources, to determine which emotional states should appear in the left branch and the right branch. For the whole speech signal, the accuracies of the 35 top level predictions distribute from 60.60% to 83.95%, with the average of 67.90% and standard deviation of 5.76%. For the glottal waveform, the accuracies distribute from 57.16% to 80.38%, with the average of 64.99% and standard deviation of 5.48%. As foreseen, the accuracies from the glottal waveform are lower than those from the original whole speech signal, while the differences are generally low, from 0.45% to 6.70%, to present the major proportion of the emotional cues by glottal waveform.

Four groups of accuracy comparison of top level are listed in Table. 3. Group 1 is the branch structure of top level with the highest accuracies. Fortunately, both types of data sources resulted into the same best structure, where the left branch contains the emotional states with relative higher arousal as moderate joy, intense joy, and intense anger (refer to Fig. 5), and the right branch contains the states with medium or lower arousal as neutral, moderate anger, moderate sadness, and intense sadness. The difference of accuracy between the two types of sources in this best case is 3.21%, slightly higher than the average of differences of all 35 cases as 2.91%. This group of left & right branches is selected as the solution of top level in the hierarchical classifier.

TABLE 3. ACCURACIES IN TOP LEVEL OF HIERARCHICAL STRUCTURE (%), AND THE DIFFERENCE OF ACCURACIES BETWEEN THE TWO TYPES OF SOURCES (%). "N" STANDS FOR NEUTRAL, "M" STANDS FOR MODERATE, "I" STANDS FOR INTENSE, "J", "A", AND "S" STAND FOR JOY, ANGER, AND SADNESS RESPECTIVELY. GROUP 1 IS THE HIGHEST, GROUP 2 IS THE LOWEST, GROUP 3 IS WITH THE MAXIMUM DIFFERENCE BETWEEN THE TWO TYPES OF SOURCE, AND GROUP 4 IS WITH THE MINIMUM DIFFERENCE.

|   | Source | Left branch | Right branch | Accu. | Diff. |
|---|--------|-------------|--------------|-------|-------|
| 1 | glottal | M-J, I-J, I-A | N, M-A, M-S, I-S | 80.38 | 3.21 |
|   | speech |  |  | 83.59 |  |
| 2 | glottal | M-J, M-A, I-S | N, I-J, I-A, M-S | 57.16 | 3.44 |
|   | speech | N, I-J, M-A | M-J, I-A, M-S, I-S | 60.60 |  |
| 3 | glottal | I-J, I-A, M-S | N, M-J, M-A, I-S | 65.63 | 6.70 |
|   | speech |  |  | 72.32 |  |
| 4 | glottal | M-A, M-S, I-S | N, M-J, I-J, I-A | 73.42 | 0.45 |
|   | speech |  |  | 73.87 |  |

TABLE 4. CONFUSION MATRICES IN TOP LEVEL (%). "L" FOR LEFT BRANCH, AND "R" FOR RIGHT BRANCH.

|   | Speech | | Glottal | |
|---|--------|---|---------|---|
|   | L | R | L | R |
| L | 83.03 | 16.97 | 78.69 | 21.31 |
| R | 15.36 | 84.64 | 18.36 | 81.64 |

The confusion matrices for the selected solution in the top level are listed in Table. 4. Both matrices are almost balanced, with a slight tendency to have more misjudgment to the right branch.

Several other possible structures in the top level are also analyzed, even though they will not be adopted in the final hierarchical architecture. Group 2 in Table. 3 is worst possible case of top level. The structure for the two types of sources are different in this case, with some common points. For both types of sources, the worst cases contain moderate joy and intense sadness in one branch, and neutral and intense joy in the other. In the valence-arousal view, both worst structures in top level involve emotional states intersect in valence or arousal axis in each branch, thus they cannot be well separated, no matter with whole speech signal or only glottal waveform.

Group 3 and 4 in Table. 3 are the cases of the highest and lowest differences between the two types of sources. In group 3, which is with the highest difference, the left branch contains intense anger, intense joy, which both have very high arousal as shown in Fig. 5, and together with moderate sadness, which has the lowest arousal; the right branch contains the other 4 emotional states with medium arousal. In group 4, which is with the lowest difference as only 0.45%, the left branch contains moderate anger, moderate sadness, and intense sadness, which all have positive valence according to Fig. 5, and the right branch contains states with zero or positive valence. These results indicate the glottal waveform could convey most emotional cues in the valence axis, while a good emotional expression in arousal axis would rely on vocal tract effect instead in a higher degree.

*2) Second level*

In the second level of the hierarchical architecture, the selected left branch and right branch (group 1 in Table. 3) are classified into sub-branches. Each branch, with 3 or 4 emotional states, gives out 3 possible dividing patterns.

The left branch, with 3 emotional states as moderate joy, intense joy, and intense anger, can be divided as any one of the 3 states against the other two. These 3 possible classifications are trained and evaluated on both sources of whole speech signal and glottal waveform, and the classification results are listed in the upper part of Table. 5. For both sources, the best cases in the second level of left branch are moderate joy, with medium arousal, against intense joy and intense anger, with high arousal. The accuracies in this case are 73.61% and 77.26% for glottal waveform and whole speech signal respectively, with the difference as 3.65%. This case is adopted in the final hierarchical in the left branch of second level. The second case, intense joy with both high arousal and high valence, against the combination of moderate joy, with medium arousal and medium valence, and intense anger, with high arousal and low valence, is more difficult to be separated. It achieved into accuracies of 64.58% and 68.19% for glottal waveform and whole speech signal. The third case, intense anger with low valence, against the two states in the joy family with medium or high valence, also resulted in relatively high accuracies, especially for whole speech signal, as 75.52%.

The right branch contains 4 emotional states as neutral, moderate anger, moderate sadness, and intense sadness. The best case in this branch is the classification of neutral & moderate anger, with medium valence, against the two states in the sadness family, with lower valence. The accuracies in this case are 67.12% and 69.04% for glottal waveform and whole speech signal, with difference of only 1.92%. This case is adopted in the final hierarchical as the right branch in the second level. The other two cases, with interleaves in valence between the left and right sub-branches, show lower accuracies, but also provide relatively lower differences between the two types of sources, as 2.38% and 3.58%.

In comparison of the structures in the two branches adopted in the second level, where the left branch mainly deals with the distinguishing in arousal, and the right branch mainly deals with the distinguishing in valence, the differences between the two types of sources are obviously lower in the right branch than in the left. Thus, we assume again that the glottal waveform can better convey emotional cues in valence axis than in arousal, as assumed in the top level.

TABLE 5. ACCURACIES IN SECOND LEVEL OF HIERARCHICAL STRUCTURE (%) AND THE DIFFERENCES (%). "M" STANDS FOR MODERATE, "I" STANDS FOR INTENSE, "J", "A" AND "S" STAND FOR JOY, ANGER, AND SADNESS RESPECTIVELY. 1, 2, 3 FOR THE LEFT BRANCH, 4, 5, 6 FOR THE RIGHT BRANCH

|   | Left | Right | Source | Accuracy | Difference |
|---|------|-------|--------|----------|------------|
| 1 | M-J | I-J, I-A | glottal | 73.61 | 3.65 |
|   |     |          | speech  | 77.26 |      |
| 2 | I-J | M-J, I-A | glottal | 64.58 | 3.61 |
|   |     |          | speech  | 68.19 |      |
| 3 | I-A | M-J, I-J | glottal | 69.92 | 5.6 |
|   |     |          | speech  | 75.52 |      |
| 4 | N, M-A | M-S, I-S | glottal | 67.12 | 1.92 |
|   |        |          | speech  | 69.04 |      |
| 5 | N, M-S | M-A, I-S | glottal | 63.25 | 2.38 |
|   |        |          | speech  | 65.63 |      |
| 6 | N, I-S | M-A, M-S | glottal | 61.13 | 3.58 |
|   |        |          | speech  | 64.71 |      |

TABLE 6. CONFUSION MATRICES IN SECOND LEVEL (%). "M" FOR MODERATE, "I" FOR INTENSE, "N", "A", "J", "S" FOR NEUTRAL, ANGER, JOY, AND SADNESS.

|            | Speech | | Glottal | |
|------------|--------|---|---------|---|
|            | M-J    | I-J & I-A | M-J | I-J & I-A |
| M-J        | 69.79  | 30.21 | 57.03 | 42.97 |
| I-J & I-A  | 19.01  | 80.99 | 18.1  | 81.9  |
|            | N & M-A | M-S & I-S | N & M-A | M-S & I-S |
| N & M-A    | 66.28 | 33.72 | 67.71 | 32.29 |
| M-S & I-S  | 28.19 | 71.81 | 33.46 | 66.54 |

The confusion matrices in the second level are listed in Table. 6. For the classification moderate joy against intense joy and intense anger, the confusion matrices are both highly unbalanced, with significant tendency to the two intense emotional states. The two sub-branches in the right branch are more balanced recognized, especially with the glottal waveform. This indicates good performance of glottal features in valence.

*3) Bottom level.*

With the selected structure from the previous two levels, the bottom level simply contains 3 sub-classifiers to distinguish two emotional states each, with no varieties. The accuracies in this level are listed in Table. 7. Two of the pairs to be classified in this level, intense joy against intense anger, and neutral against moderate anger, have similar arousal levels within the pairs, while different valence levels, get low difference between sources of whole speech signal and glottal waveform, as 3.13% and 2.02%. The other pair, moderate sadness and intense sadness, with different levels in both arousal and valence, shows higher difference between the two types of sources, as 4.1%. The assumption that the glottal features express better valence cues than arousal cues is confirmed again.

TABLE 7. ACCURACIES IN BOTTOM LEVEL OF HIERARCHICAL STRUCTURE (%) AND THE DIFFERENCE. "N" STANDS FOR NEUTRAL, "M" STANDS FOR MODERATE, "I" STANDS FOR INTENSE, "J", "A", AND "S" STAND FOR JOY, ANGER, AND SADNESS RESPECTIVELY.

| Left | Right | Source | Accuracy | Difference |
|---|---|---|---|---|
| I-J | I-A | glottal | 65.82 | 3.13 |
|     |     | speech  | 68.95 |      |
| N   | M-A | glottal | 62.17 | 2.02 |
|     |     | speech  | 64.19 |      |
| M-S | I-S | glottal | 71.03 | 4.1  |
|     |     | speech  | 75.13 |      |

TABLE 8. CONFUSION MATRICES IN BOTTOM LEVEL (%). "M" FOR MODERATE, "I" FOR INTENSE, "N", "A", "J", "S" FOR NEUTRAL, ANGER, JOY, AND SADNESS.

|     | Speech |       | Glottal |       |
|-----|--------|-------|---------|-------|
|     | I-J    | I-A   | I-J     | I-A   |
| I-J | 73.18  | 26.82 | 65.49   | 34.51 |
| I-A | 35.29  | 64.71 | 33.85   | 66.15 |
|     | N      | M-A   | N       | M-A   |
| N   | 67.71  | 32.29 | 60.94   | 39.06 |
| M-A | 39.32  | 60.68 | 36.59   | 63.41 |
|     | M-S    | I-S   | M-S     | I-S   |
| M-S | 80.99  | 19.01 | 72.14   | 27.86 |
| I-S | 30.73  | 69.27 | 30.08   | 69.92 |

The confusion matrices in the bottom level are listed in Table. 8. In this level, the glottal waveform present more balanced pattern than speech signal in all 3 cases of this level.

Upon the generation process, the final hierarchical architecture is shown in Fig. 11. The overall classification accuracies of the hierarchical on the 7 emotional states are 39.98% and 44.76% for glottal waveform and speech signal respectively, with difference of 4.78%. These accuracies seem to be quite low, but are still several times as the chance level for 7 classes, $1/7 \approx 14.29\%$. For the major purpose of this work, to find out in what degree that glottal stimulation can convey the emotional cues in speech signal, the consistent assumptions drawn from the experimental results are already rather clear. Better machine learning methods, *e.g.*, deep learning approaches, would lead to higher accuracies in classification in our future work.

### C. Experiment II: classification on individual emotional state pairs

In the previous experiment, the accuracies are generally low, because in the top level, emotional states that relatively similar while still different to each other are put together in the left or right branch, to interfere the separation; in the second or bottom level, the task changes to distinguish emotional states that have strong common properties. In this experiment, we will not group any emotional states together, while only study the pairs of individual emotional states.

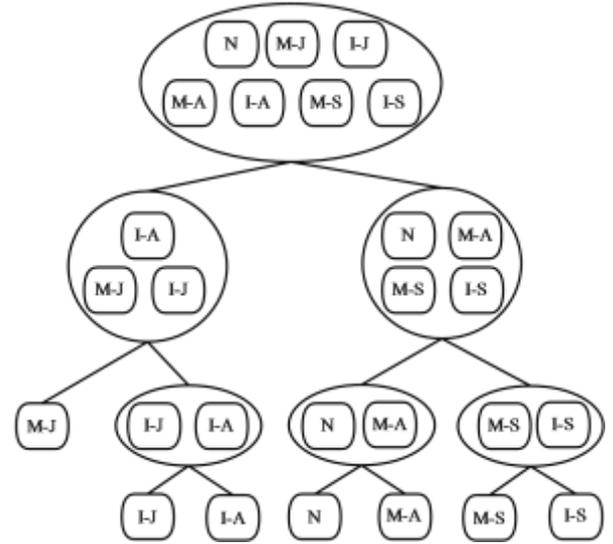

Figure 11. Hierarchical architecture generated according to binary classification accuracy. "N" stands for neutral, "M" stands for moderate, "I" stands for intense, "J", "A", and "S" stand for joy, anger, and sadness respectively.

$C_7^2 = 21$ pairs are possible for the 7 emotional states in the MES-P dataset. Classifications on all the pairs with SVMs are performed with features extracted from original whole speech signal or glottal waveform. The accuracies and the differences between the two types of sources are listed in Table. 9.

We first analyze the accuracies. The best classified emotional pair is intense anger against moderate sadness, with the accuracy up to 94.73% for whole speech signal and 92.45% for glottal waveform, followed by the pairs: intense joy vs. moderate sadness (94.53% and 90.10%), neutral vs. intense anger (94.40% and 88.93%), neutral vs. intense joy (94.14% and 88.93%), intense anger vs. intense sadness (91.21% and 84.51%). All these pairs with high accuracy (> 90% for whole speech signal) involve one of the two states of very high arousal, i.e. intense anger or intense joy. The classification accuracy between these two states, however, is as low as 68.95% for whole speech signal and 65.82% for glottal waveform, which is among the worst pairs. Take the accuracies from glottal waveform as reference, 2 pairs get accuracy above 90%, 5 pairs below 90% but above 80%, 9 pairs below 80% but above 70%, and only 5 pairs below 70%. In the hierarchical in the previous experiment, the accuracies in the 3 sub-classifiers in the bottom level are ranked in the 21 pairs as 16, 19, and 20, all among the lowest pairs, because the emotional states with significant divergence have been separated in previous levels.

The difference in performances from speech signal and from glottal waveform is the other point to be studied in this experiment. These differences are illustrated in Fig. 12 with different gray scales, darker grids correspond to pairs with higher difference. There are 3 pairs exhibit differences higher than 7%: moderate joy vs. intense joy, moderate joy vs.

TABLE 9. ACCURACIES IN EMOTIONAL STATE PAIRS (%) AND THE DIFFERENCE. "N" STANDS FOR NEUTRAL, "M" STANDS FOR MODERATE, "I" STANDS FOR INTENSE, "J", "A", AND "S" STAND FOR JOY, ANGER, AND SADNESS RESPECTIVELY.

|   | Source | M-J | I-J | M-A | I-A | M-S | I-S |
|---|---|---|---|---|---|---|---|
| N | glottal | 74.15 | 88.93 | 62.17 | 88.93 | 68.03 | 77.28 |
|   | speech | 76.37 | 94.14 | 64.19 | 94.4 | 68.88 | 79.49 |
|   | difference | 2.21 | 5.21 | 2.02 | 5.47 | 0.85 | 2.21 |
| M-J | glottal |   | 66.47 | 72.92 | 71.61 | 78.71 | 78.32 |
|   | speech | - | 74.28 | 74.48 | 81.64 | 80.21 | 81.32 |
|   | difference |   | 7.81 | 1.56 | 10.03 | 1.5 | 2.99 |
| I-J | glottal |   |   | 84.11 | 65.82 | 90.1 | 79.95 |
|   | speech | - | - | 88.74 | 68.95 | 94.53 | 89.58 |
|   | difference |   |   | 4.62 | 3.13 | 4.43 | 9.64 |
| M-A | glottal |   |   |   | 84.38 | 62.11 | 71.61 |
|   | speech | - | - | - | 87.5 | 68.62 | 77.93 |
|   | difference |   |   |   | 3.13 | 6.51 | 6.32 |
| I-A | glottal |   |   |   |   | 92.45 | 84.51 |
|   | speech | - | - | - | - | 94.73 | 91.21 |
|   | difference |   |   |   |   | 2.28 | 6.71 |
| M-S | glottal |   |   |   |   |   | 71.03 |
|   | speech | - | - | - | - | - | 75.13 |
|   | difference |   |   |   |   |   | 4.1 |

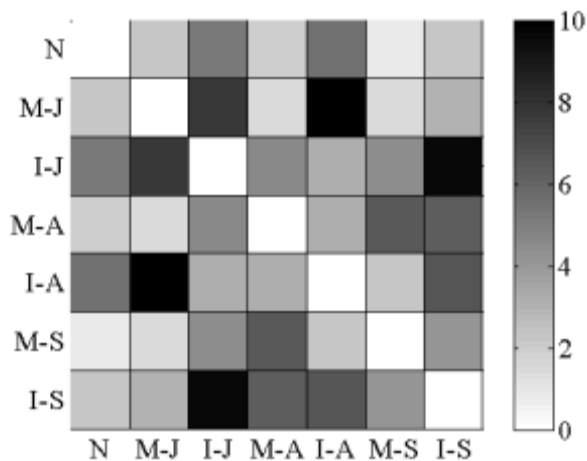

Figure 12. Difference of accuracy in pair-wise classification between the two types of source. "N" stands for neutral, "M" stands for moderate, "I" stands for intense, "J", "A", and "S" stand for joy, anger, and sadness respectively.

intense anger, intense joy vs. intense sadness. All these 3 pairs have obvious distances in both arousal and valence axis. The pair with the lowest difference is neutral vs. moderate sadness (0.85%). These two states are relatively close to each other in arousal. There are also other pairs with close arousal levels, such as neutral vs. moderate joy, neutral vs. moderate anger, moderate joy vs. moderate anger, the differences are all low (1.56% to 2.21%). This also accords with the findings in the previous experiment.

Upon the comparative experiments in this section, we assume that glottal waveform can convey a majority of emotional cues expressed in speech, by the classification accuracies from glottal waveform that are generally only slightly lower than those from speech signal with leave-one-speaker-out SVMs. The differences between the two types of sources vary from 0.45% to 10.03%. Most cases of our experiments indicate that to distinguish emotional state(s) with similar arousal levels is more likely to get lower differences between the two types of sources, while emotional state(s) with different arousal tend to lead to higher differences. Thus, we can draw the conclusion that more emotional cues can be convey by only glottal waveform in valence axis than in arousal axis.

## V. CONCLUSION

A comparative investigation on the contribution of the glottal waveform to the speech emotion expressing is stated in this work.

We first compared the glottal waveform extracted from a small number of speech samples in different emotional states, confirmed that there exist significant difference among different emotional states, in both time domain and frequency domain of glottal waveform.

Automatic machine learning based comparison is done in this work to find out in what degree that the glottal waveform can express the emotion in human vocal communications. Identical hierarchical classification architectures are generated according in a performance-driven manner for both types of sources as original whole speech signal and extracted glottal waveform. In the generated hierarchical classifier architecture, the difference in accuracy between the two types of sources varies from 1.92% to 5.6% in the sub-classifiers. The low differences can lead to the conclusion that a majority of emotional cues are contained in the stimulation source of speech, i.e. the glottal waveform.

Another regular pattern found in the experimental results is that the difference in accuracy between speech signal and glottal waveform is generally higher for emotional states with larger distances in arousal, lower for states with similar arousal levels, regardless of the distance in valence. The worse performance of the glottal waveform in arousal axis lead to conclusion that the glottal waveform can carry more emotional cues in valence, and less in arousal.

The influence of vocal tract effect to emotion, which is not taken into account in this work but also essential, will be investigated in our future work. In addition to the common exhibition ability of emotional cue by the vocal tract filtering and glottal stimulation, we are interested in its unique emotion expressing properties, hopefully to compensate the deficiency in arousal by glottal waveform.